# SYNTHESIS OF CARBON NANOTUBES

Kalpana Awasthi, Anchal Srivastava and **O.N. Srivastava***

Physics Department, Banaras Hindu University, Varanasi-221 005, INDIA

---------------

Corresponding author : Phone & Fax : +91-542-2368468 / 2307307

E-mail : hepons@yahoo.com

# CONTENTS

## 1. INTRODUCTION

Until 1985 it was generally believed that solid elemental carbon occurs in two different crystalline phases: diamond and graphite. Diamond is in thermodynamic equilibrium at very high temperatures and pressures; it occurs, nevertheless, as a metastable phase under atmospheric pressure and at room temperature. In the structure of diamond [Fig. 1(a)] each carbon atom is tetrahedrally surrounded by four $sp^3$ covalently bonded carbon atoms. The resulting spatial network of carbon is built on a cubic face centred lattice. The structure of graphite [Fig. 1(b)] consists of graphene layers in which the $sp^2$ bonded carbon atoms form a planar hexagonal honeycomb arrangement. The bonding of carbon atoms in a graphene plane is very strong (covalent bonds), whereas the bonding between two graphene layers is weak (Vander Waals bonds). In 1985 an important breakthrough in carbon research was realized by the work of Kroto et al.,[1] which resulted in the discovery of large family of all carbon molecules, called 'fullerenes'. They can be crystallized as molecular crystals, which are thus a third form of crystalline elemental carbon. The fullerenes are closed cage carbon molecules with the carbon atoms tiling spherical or nearly spherical surfaces, the best known example being $C_{60}$ with a truncated icosahedral structure formed by 12 pentagonal rings and 20 hexagonal rings [Fig.1(c)]. The coordination at every carbon atom in fullerenes is not planar but rather slightly pyramidalized, with some $sp^3$ bonding present in the essentially $sp^2$ carbons. $C_{60}$ molecular structure shows that every pentagon of $C_{60}$ is surrounded by five hexagons. The key feature is the presence of five-membered rings, which provide the curvature necessary for forming a closed-cage structure. In 1990, Kratschmer et al.[2] found that the soot produced by arcing graphite electrodes contained $C_{60}$ and other fullerenes. It was the ability to generate fullerenes in gram-scale quantities in the laboratory, using a relatively simple apparatus that gave rise to intense research activity on these molecules and caused a renaissance in the study of carbon.

Carbon nanotubes (CNTs) were discovered by S. Iijima, [3] who was looking for new carbon structures, in the deposit formed on graphite cathode surfaces during the electric-arc evaporation (or discharge) that is commonly employed to produce fullerene soot. The CNTs, also known as tubular fullerenes, are cylindrical graphene sheets of $sp^2$ bonded carbon atoms. These nanotubes are concentric graphitic cylinders closed at either end due to the presence of five-membered rings. The CNTs can be multiwalled with a central tube of nanometric diameter surrounded by graphitic layers separated by ~0.34nm [Fig.1 (d)]. Unlike the multi-walled carbon nanotubes (MWNTs), in single-walled carbon nanotubes (SWNTs) there is only the tube and no graphitic layers i.e. SWNTs consist of singular graphene cylindrical walls. In 1999, Rode et al., [4,5] prepared a new form of carbon, a low-density cluster assembled carbon nanofoam. Carbon nanofoam has been prepared by a high-repetition-rate, high-power laser ablation of glassy carbon in Ar atmosphere. The nanofoam possesses a fractal-like structure consisting of carbon clusters with an average diameter of 6-9 nm randomly interconnected into a web-like foam.The nanofoam is the first form of pure carbon to display ferromagnetism albeit temporary, at room temperature .[6]

Ever since, the discovery of CNTs, several ways of preparing them has been explored. The CNTs have been synthesized by various methods e.g. electric arc discharge, laser evaporation and chemical vapor deposition.[7-9] These methods are very useful and are of widespread importance. The CNTs can be inert and can have a high aspect ratio, high tensile strength, low mass density, high heat conductivity, large surface area and versatile electronic behavior including high electron conductivity. The combination of these properties makes them ideal candidates for a large number of applications provided their cost is sufficiently low. Electronically, CNTs are expected to behave as ideal one-dimensional 'Quantum Wires" with either semiconducting or metallic behavior, depending upon the diameter and orientation of tube axis (parallel or perpendicular to the C-C bond). [10-12] Collins et al. [13] reported an experimentally functioning carbon nanodevice (diode) based on nanotube having, a variety of structure cells. Room temperature transistor based on a single CNT has been reported which marks an important step towards molecular electronics. [14,15] Recent developments have focused considerable media attention on nanotube nanoelectronic application with crossed SWNTs, three and four terminal electronic devices have been made, as well as a nonvolatile memory that functions like an electromechanical relay.[16, 17] Integrated CNT devices involving two CNT transistors have been reported, providing visions of large-scale integration. [18] The small diameter of CNTs is very favorable for field emission- the process by which a device emits electrons when an electric field or voltage is applied to it. Aligned CNTs are considered to be ideal for the purpose because of their high packing density and hence for use as high brightness field emitters. The use of CNTs as field emitters was first proposed by de Heer et al. in 1995. [19] A current density of 0.1mA/cm$^2$ was observed for voltages as low as 200V. Lee et al. [20] have shown that aligned CNT bundles exhibit a high emission current density of around 2.9mA/cm$^2$ at 3.7V/μm.

The exceptional mechanical properties and low weight of nanotubes make them potential filling materials in polymer composites. Nanotubes can improve the strength and stiffness of a polymer, as well as add multifucntionality (such as electrical conductivity) to polymer based composite systems. [21-25] The CNTs should be ideal reinforcing fibres for composites due to their aspect ratio and high in axis strength. [26] Nanotubes can also mechanically deflect under electric stimulation (e.g. due to charge induced on the nanotubes) and this opens up applications such as cantilevers or actuators. [25,27,28] The CNT containing ceramic-matrix composites are a bit more frequently studied, most efforts made to obtain tougher ceramics. [29] It has also been suggested that nanotubes might be used as membrane material for batteries and fuel cells, anode for lithium ion batteries, capacitor and chemical sensors/ filters.[30-32] The high electrical conductivity and relative inertness of nanotubes make them potential candidates as electrodes in electrochemical reactions too. [33] The large surface area of nanotubes, both inside and outside, can be usefully employed to support reactant particles in catalytic conversion reactions. [34,35] Nanotube tips for probe microscopies have also been designed and nanotubes also possess hydrogen storage capability. [36-38] Many of these proposed applications will require large-scale synthesis of CNTs in high purity. Recent reviews summarize possible technological applications with focus on the properties and catalytic synthesis of CNTs. [39-43]. Our review begins with an overview of the different methods of preparation of CNTs. Special CNTs configurations, such as nanocoils, nanohorns, bamboo-shaped CNTs and carbon cylinder made up from CNT will also be discussed.

## 2. DIFFERENT SYNTHESIS TECHNIQUES FOR CARBON NANOTUBES

There are many techniques used to produce MWNTs or SWNTs. Methods such as electric arc discharge, laser vaporization and chemical vapour deposition techniques are well established to produce a wide variety of CNTs. These methods are described in following sections.

### 2.1 Electric-Arc Discharge

Carbon nanotubes (CNTs) are commonly prepared by striking an arc between graphite electrodes in an inert atmosphere (argon or helium), the process that also produces carbon soot containing fullerene molecules. [1] The carbon arc provides a convenient and traditional tool for generating the high temperatures needed to vaporize carbon atoms into a plasma (>3000$^{o}$C). [7,44,45] The yield of CNTs depends on the stability of the plasma formed between the electrodes, the current density, inert gas pressure and cooling of electrodes and chamber. [7,45] Among the various inert gases, helium (He) gives the best results, probably due to its high ionization potential. [46] The well-cooled electrodes and arc chamber help to maximize the nanotube yield in the arc growth process. For the MWNTs production, the conditions are optimized so that during the arc evaporation, the amount of soot production is minimized and 75% of the evaporated carbon from a pure graphite anode is made to deposit onto the facing graphite cathode surface. The arc deposit consists of a hard gray outer shell made of pyrolitic graphite and an interior made of a soft black powder containing about two thirds CNTs and one third graphitic nanoparticles. The optimized synthesis conditions were at 20-25V, 50-100Amp. d.c. (direct current) and the helium pressure maintained at 500 torr. Arc discharge is a simple process, and it is the method to obtain structurally excellent high quality CNTs. However, conventional arc discharge is a discontinuous and unstable process and it cannot produce the large quantity of CNTs. The CNTs are produced on the cathode surface and the electrode spacing is not constant, so the current flow is not uniform and the electric fields are non-homogeneous. As the result, the density of carbon vapor and the temperature distribution is non-uniform and carbon nanoparticles and impurities always co-exist with nanotubes. In order to solve this problem, many efforts to generate the stable and high efficient discharge have been made, and many studies to understand the growth mechanism of nanotube have been done. [47,48] Lee et al. [49] prepared CNTs in large amount by plasma rotating arc discharge method. In this process the graphite anode is rotated at a high velocity for the synthesis of CNTs. The rotation of the anode distributes the microdischarges uniformly and generates stable plasma. The centrifugal force by the rotation generates the turbulence and accelerates carbon vapor perpendicular to the anode. It is not condensed at the cathode surface but collected on the graphite collector that was placed at the periphery of the plasma. The nanotube yield increases as the rotation speed of the anode increases and the collector becomes closer to the plasma. The reason for this is because two conditions are optimized. One is high density of carbon vapor that is created by uniform and high temperature plasma for nucleation and the other is the sufficient temperature of collectors for nanotube growth. The plasma rotating electrode process is continuous process of the stable discharge and it is expected to perform the mass production of high quality nanotubes. The CNTs have been produced by using plasma arc jets and in large quantities by optimizing the quenching process in an arc between a graphite anode and a cooled copper electrode .[50,51]

Ishigami et al. [52] have reported the simplified arc method for the continuous synthesis of CNTs. This method requires only a d.c. power supply, graphite electrodes and a container of liquid nitrogen, there is no need for pumps, seals, water-cooled vacuum chambers, or purge-gas handling systems, which are necessary for the production of CNTs via conventional arc discharge. They reported that the reaction can run in a continuous fashion and can be scaled up for industrial applications with a CNT yield comparable to that of an optimized conventional arc reaction. In general, the nanotubes were of high quality, had four to eight layers and had long and straight parallel walls with only occasional surface contamination. In fact, the tubes grown using the liquid-nitrogen method appear to have consistently cleaner surfaces than tubes grown using other methods. The tubes are composed only of carbon and show no evidence of nitrogen incorporation. This carbon arc nanotube synthesis method eliminates nearly all the complex and expensive equipment associated with conventional nanotube growth techniques.

In the arc discharge method synthesis of MWNTs require no catalyst, catalyst species are however, necessary for the growth of SWNTs. The first report on the production of SWNTs was by Ijima and Ichihashi.[53] These authors produced SWNTs material by arcing Fe-graphite electrode in a methane argon atmosphere. In this case, a hole is made in the graphite anode, which is filled with a composite mixture of metal and graphite powders, while the cathode is pure graphite. The catalyst used to prepare isolated SWNTs include transition metals such as Fe, Co, Ni and rare earth metals such as Y and Gd, [53-57] whereas composite catalyst such as Fe/Ni and Co/Ni have been used to synthesize ropes (bundles) of SWNTs.[58] In these experiments, the tubes exhibited an average diameter of 1.2 nm. Saito et al. [59] compared SWNTs produced by using different catalyst and found that a Co or a Fe/Ni bimetallic catalyst gives rise to tubes forming a highway-junction pattern. Ni catalyst yield long and thin tubes radially growing from the metal particles. High yield of SWNTs has been synthesized by d.c. arc discharge under low pressure of helium gas (100 torr) with a small amount of a mixture of Ni, Fe and graphite powders. [60] In addition, they introduced sulfur promoter to improve the yield, which gave rise to again the highest yield at low gas pressure. SWNTs were also prepared by using various oxides ($Y_2O_3$, $La_2O_3$, $CeO_2$) as catalysts. [61] For Ni-Y-graphite mixtures, Journet et al. [62] found that high yields (~80%) of SWNTs (average diameter of ~1.4 nm) can be produced. This Ni-Y mixture is now used worldwide for production of SWNTs in high yield. Shi et al. [63] reported the large-scale production of SWNTs under the arc conditions of 40~50A d.c. and helium pressure of 500 or 700 torr by using a graphite rod with a hole filled with the powder of a mixture of Y-Ni alloy and graphite or $CaC_2$-Ni and Ni as anode.

In order to raise the production of SWNTs, Liu et al. [64] used a semi-continuous hydrogen arc discharge method, they obtained ~2g/h SWNT bundles, the production of SWNTs in the ropes about 30%, the diameter of SWNTs is about 1.72 nm. Ando et al. [65] developed a d.c. arc plasma jet method and the highest yield was 1.24g/min and the purity about 50%. Takizawa et al. [66] studied the effect of environmental temperature for synthesizing SWNTs by arc vaporization. Recently, large scale and high purity of SWNTs has been synthesized by an arc discharge under controlled temperatures with multi-metal catalysts of Fe-Ni-Mg in He atmosphere.[67] The temperature strongly affects the yield of SWNTs and it increases as increasing temperature. The SWNT bundles with diameters 7-20 nm and the production of 45.3 g/h were prepared at 600$^{o}$C.

The arc method usually involves high-purity graphite electrodes, metal powders (only for producing SWNTs), and high-purity He and Ar gases; thus the cost associated with the production of SWNTs and MWNTs are high. Although the crystallinity of the material is also high, there is no control over dimensions (length and diameter) of the tubes. Unfortunately, by-products such as polyhedral graphite particles (in the case of MWNTs), encapsulated metal particles (for SWNTs), and amorphous carbon are also formed.

## 2.2 Laser Vaporization

An efficient route for the synthesis of bundles of SWNTs with a narrow distribution is the laser evaporation technique. In this method, a piece of graphite target is vaporized by laser irradiation under high temperature in an inert atmosphere. MWNTs were found when a pure graphite target was used.[8] The quality and yield of these products have been found to depend on the reaction temperature. The best quality is obtained at $1200^{\circ}$C reaction temperature. At lower temperatures, the structure quality deceases and the CNTs start presenting many defects. As soon as small quantities (few percents or less) of transition metals (Ni, Co) playing the role of catalysts are incorporated into the graphite pellet, products yielded undergo significant modifications and SWNTs are formed instead of MWNTs. The yield of SWNTs strongly depends on the type of metal catalyst used and is seen to increase with furnace temperature, among other factors. A high yield with about 50% conversion of transition-metal/graphite composite rods to SWNTs was reported in the condensing vapor in a heated flow tube (operating at $1200^{\circ}$C).[68] Depending on the metal catalyst used, the yield on the mono or bimetal catalysts are ordered as follows: Ni>Co>Pt>Cu or Nb and Co/Ni, Co/Pt>Ni/Pt>Co/Cu, respectively. A remarkably high nanotube yield of 50% on a Co/Ni run might have resulted from some uncertainty in the process of catalyst preparation and pretreatment.[69]

In 1996, dual pulsed laser vaporization was used to optimize the laser oven method further to produce SWNT yields of 70%.[69] Samples were prepared by laser vaporization of graphite rods with 1.2 at% of a 50:50 mixture of Co and Ni powder at $1200^{\circ}$C in flowing argon at 500 torr, followed by heat treatment in vacuum at $1000^{\circ}$C to sublimate the $C_{60}$ and other smaller fullerenes. In this method, the amount of carbon deposited as soot is minimized by the use of two successive laser pulses: the first to ablate the carbon-metal mixture and the second to break up the larger ablated particles and feed them into the growing nanotube structures. These SWNTs were nearly uniform in diameter and self-assemble into ropes (bundles) which consisting of 100 to 500 tubes in a two-dimensional triangular lattice. It is also possible to produce

SWNTs using a $CO_2$-laser focused on a graphite-metal target in the absence of an oven.[70,71] In this context, Ar and $N_2$ were determined to be the best atmospheres to generate SWNT bundles, whereas He produced only small amounts of CNTs. Similarly, Dillon and coworkers [72] noticed that the diameter of the tubes depends upon the laser power. Other laser experiments revealed that porous targets of graphite-Co-Ni (nitrate) yield twice as much SWNT material from the standard metal carbon target.[69, 73] Recently, Eklund et al. [74] reported that ultrafast (subpicosecond) laser pulses are able to produce large amounts of SWNTs.

Unfortunately, the laser technique is not economically advantageous because the process involves high-purity graphite rods, the laser powers required are high (in some cases two laser beams are required), and the amount of CNTs that can be produced per day is not as high as arc discharge method.

**2.3 Chemical Vapor Deposition**

Chemical vapor deposition (CVD) is one of the most popular thin film deposition method. CVD is very different from the other two common methods used for CNT production, namely electric arc discharge and laser vaporization.[7, 62,69] Arc discharge and laser vaporization can be classified as high temperature (>3000K) and short time reaction (μs-ms) techniques, whereas catalytic CVD is a medium temperature (700-1473K) and long time reaction (typically minutes to hours) technique. The main technological drawbacks with arc discharge and laser vaporization are that the CNTs are produced as stand alone on their own.[75,76] The CNTs do not grow on a conventional or patterned substrate.

Earlier most CVD-grown CNTs were "spaghetti-like" and defective, but the potential of the technique to satisfy technological requirements was recognized. From 1998 onward, substantial and rapid progress has been made in the development of CVD to establish it as a highly controlled technology for the production of CNTs. Today, it is possible to fabricate high quality multi-walled carbon nantoubes (MWNTs) and single-walled carbon nantoubes (SWNTs) directly onto substrates or in bulk as a raw material.[9, 77-79] A major advantage of CVD is that the CNTs can be used directly without further purification unless the catalyst particle is required to be removed. In CVD method, CNTs are grown by decomposing an organic gas over a substrate covered with metal catalyst particles. Some CVD methods are reported, such as: thermal CVD, plasma enhanced CVD and catalytic pyrolysis of hydrocarbon.

### 2.3.1 Thermal Chemical Vapor Deposition (Decomposition of Hydrocarbon Gas on Metal Catalysts)

A thermal CVD reactor is simple and inexpensive to construct and consists of a quartz tube enclosed in a furnace. The substrate material may be Si, mica, silica quartz or alumina. The nature and yield of the deposit obtained in the reaction are controlled by varying different parameters such as the nature of the catalytic metals and their supports, the hydrocarbon sources, the gas flows, the reaction temperature, and the reaction time, etc. By selecting proper conditions, both the physical (e.g. length, shape, diameter) and chemical properties (e.g. defects, graphitization) of CNTs can be designed in advance. Most of the thermal CVD methods used to grow MWNTs use acetylene ($C_2H_2$) or ethylene ($C_2H_4$) gas as the carbon feedstock and Fe, Ni or Co nanoparticles as the catalyst. The growth temperature is typically in the range 500-900$^oC$. At these temperatures, the carbon atoms dissolve in the metal nanoparticles, which eventually become saturated. The carbon then precipitates to form CNTs, the diameters of which are determined by the sizes of the metal particles, which work as catalyst. When other elements (e.g. Cu, Cr, Mn) are used, only a negligible amount of CNTs is formed. The recently reported experimental results are shown in Table 1. Fonseca et al. [80] produced large amounts of MWNTs by catalytic deposition of $C_2H_2$ over Co and Fe catalysts supported on silica or zeolite. They investigated the influence of various parameters, such as catalysts preparation, the nature of the support, the size of active metal particles and the reaction conditions on CNT formation. Clay supports in the form of Co/Clay and Fe/Clay were found to have low activity in the formation of CNTs (~10-12% in 30min). [81] Instead of single metal catalysts, Co/Mo, Co/V and Co/Fe mixtures supported by either zeolite or alumina have also been used catalysts to decompose $C_2H_2$ to produce MWNTs.[82] Co is necessary for the formation of good quality MWNTs, whereas Fe is actively responsible for the thickness of the tubes. Mixtures of Co and V in equal amounts produced the highest yields and thinest tubes compared to the other compositions.

The choice of the carbon feedstock also affects the growth of CNTs. Baker and Harris [83] reported that unsaturated hydrocarbons such as $C_2H_2$ had much higher yields and higher deposition rates than saturated gases. They also observed that saturated carbon gases tend to produce highly graphitized filaments with fewer walls compared with unsaturated gases. Thus, hydrocarbons such as $CH_4$ and CO are commonly used for SWNT growth [9,84-89], whereas hydrocarbons such as $C_2H_2$, $C_2H_4$ and $C_6H_6$, which are unsaturated and thus have high carbon content, are typically used for MWNT growth .[80,82, 90-94] Dai et al. [84] first synthesized SWNTs by CVD on a $Mo/Al_2O_3$ catalyst with carbon monoxide (CO) as the carbon feedstock at 1200$^{o}$C, but the yield was very low. They found that Mo particles are localized at the tips of the SWNTs (1-5 nm). Peigney et al. [85] then grew a mixture of SWNTs and MWNTs from the decomposition of $CH_4$ on $Fe/Al_2O_3$ nanoparticles. The optimized catalysts consisted of Fe/Mo bimetallic species supported on a novel silica-alumina composite material and produce ~42wt% SWNTs that consist of individual and bundles of SWNTs that are free of defects and amorphous carbon coating. Kitiyanan et al. [86] tailored Co/ Mo catalysts to maximize the selectivity to SWNTs to minimize the subsequent purification steps. They demonstrated "controlled production" of SWNTs, based on a simple quantification method-temperature programmed oxidation (TPO)- that; allow them to conduct systematic screening of catalyst formulations and operation conditions. Colomer et al. [87] obtained a product with high yields (70-80%) of SWNTs by catalytic decomposition of $H_2/CH_4$ mixture over well-dispersed metal particles (Co, Ni, Fe) on MgO at 1000$^{o}$C. The MgO based support is easily removed by mild acidic treatment that does not damage the CNTs. The synthesis of composite powders that contain well-dispersed CNTs via selective reduction in $H_2/CH_4$ of oxide solid solutions has been reported between a non-reducible oxide such as $Al_2O_3$ or $MgAl_2O_4$ and one or more transition metal oxide.[85,95-97] The decomposition of $CH_4$ over the freshly formed nanoparticles prevents further growth and results in a very high proportion of SWNTs and less MWNTs. The mixtures of SWNTs and double-walled CNTs (DWNTs) also were synthesized in the same way from a $Mg_{0.9}Co_{0.1}O$ solid solution that was prepared by combustion synthesis of a stoichiometric mixture of metal nitrates and urea.[98] The DWNTs, consist of two concentric graphene cylinders, a structure that is intermediate between single and multi-walled CNTs. Flahaut et al. [99] also have prepared gram - scale amounts of clean double-walled carbon nanotubes (DWNTs), with a good selectivity and with low residual catalyst content. In their experiment the fuel (urea) was replaced by the equivalent amount of citric acid. They have shown that the addition of a very small amount of Mo is very efficient at increasing the yield of CNTs and that it also increases the selectivity towards DWNTs. Cumings et al. [100] have synthesized DWNTs by catalytic decomposition of $CH_4$ over Fe/ $Al_2O_3$ mixture. They utilized fumed alumina as a catalyst support material and deposited iron salt onto the support from a methanol solution. More recently, Endo et al. [101] reported the formation of DWNTs via thermolytic processes involving Mo and Fe catalysts in conjunction with $CH_4$.

The size of the catalyst is probably the most important parameter for the nucleation of SWNTs. Li et al. [102] prepared catalyst nanoparticles of uniform diameters (between 3 to 14 nm) by thermal decomposition of metal carbonyl complexes using a mixture of long-chain carboxylic acid and long-chain amine as protective agents. Their results indicate that the upper limit for SWNT growth occurred at catalyst sizes between 4 and 8 nm. They also grew SWNTs from discrete catalytic nanoparticles of various sizes.[103] Discrete nanoparticles were prepared by placing a controllable number of metal atoms into the cores of apoferritin. Smaller nanoparticles (≤1.8nm) were more active in producing SWNTs, while nanoparticles with diameters of ~7nm did not show SWNTs growth. Cheung et al. [104] prepared monodispersed nanoclusters of Fe with diameters of 3, 9 and 13 nm. After growth using $C_2H_4$, single-walled and double walled nanotubes were nucleated from the 3 and 9 nm diameter nanoclusters, whereas only MWNTs were observed from the 13 nm nanoclusters. These works clearly suggest that SWNTs are favored when the catalyst particle size is ~5nm or less. Recently, Zheng et al. [105] reported the synthesis of 4-cm long individual SWNTs at a high growth rate of 11 μm/sec by catalytic CVD. The ultralong SWNTs were synthesized by Fe-catalyzed decomposition of ethanol.

Supported catalyst (Fe, Co, Ni) that contain either a single metal or a mixture of metals seem to induce the growth of isolated SWNTs or SWNTs bundles, respectively, in the $C_2H_4$ atmosphere. [106] For the growth of CNTs, $Fe/SiO_2$ shows the highest activity in the decomposition of unsaturated compounds. The activity of Co/Y is much lower, but the quality of CNTs is the best, displaying almost perfect graphitization. [107] The amorphous carbon is deposited due to the decomposition of the hydrocarbon gases used. Some attempts have been made to prevent the formation of amorphous carbon such as the addition of hydrogen (which etches amorphous carbon) into the deposition process together with the hydrocarbon gas or by performing growth at very high gas pressures (> 1atm), which would inhibit the C species from sticking to the substrates, and therefore prevent the accumulation of carbon.[9, 108] Franklin et al. [108] found that for clean SWNT growth at 900°C, the optimum flow of hydrogen was between 100 and 150 ml/min in a predominantly $CH_4$ flow (~1500 ml/min). The flow of hydrogen had to be increased to 200 ml/min to maintain clean growth if a temperature of 950°C was used. In the absence of hydrogen flow, the $CH_4$ was found to decompose and form amorphous carbon deposits all over the substrate. Recently, Lacerda et al. [109] have demonstrated a simple way to deposit SWNTs by CVD without the co-deposition of unwanted amorphous carbon. Using a triple-layer thin film of Al/Fe/Mo (with Fe as a catalyst) on an oxidized Si substrate, the sample was exposed to a single short burst (5 sec) of $C_2H_2$ at 1000°C. They believe that the high temperature is responsible for the high crystallinity/straightness of the nanotubes and the rapid growth process allows us to achieve a clean amorphous carbon free deposition, which is important for SWNT device fabrication.

Maruyama et al.[110] have developed a method of producing high-quality SWNTs that uses alcohol as a carbon feedstock. This method performs well not only for the mass production of SWNTs, but also for the direct synthesis of SWNTs on non-metallic substrates such as silicon and quartz. For the mass production of SWNTs, this method can achieve a yield of >40wt% over the initial weight of catalytic powder, that is, USY-zeolite supporting 5wt% catalytic metals, within the CVD reaction time of 120 min. Fig. 2 shows a typical transmission electron microscopy (TEM) image of as-grown SWNTs. Typical as-grown SWNTs synthesized by the alcohol catalytic CVD method show no metal particles among the bundles of SWNTs and almost no amorphous carbons adhering on the tube walls. Furthermore, no MWNTs were observed in the sample shown in Fig. 2. This method does not use a conventional deposition/sputtering technique in the mounting of catalytic metals on the surface of substrate, but rather a unique, easy, and cost less liquid – based dip-coat technique in which a piece of the substrate was vertically drawn up from the metal acetate solution at a constant speed. Fig. 3 shows the scanning electron microscope (SEM) image of a quartz substrate (thickness ~0.5 mm, both sides optically polished) taken from a tilted angle including a broken cross-section of the substrate. The substrate was blackened and a uniform mat of SWNTs with a thickness of a few hundred nanometers was formed on both sides of the quartz substrate.

### 2.3.2 Plasma Enhanced Chemical Vapor Deposition

Plasma-enhanced chemical vapor deposition (PECVD) is a promising up-coming growth technique for the selective positioning and vertical alignment of CNTs. Vertical alignment is important applications. This is very useful in field emitters, which are currently being considered for use in flat panel displays. The conventional wisdom in choosing plasma processing is that the precursor is dissociated by highly energetic electrons and as a result, the substrate temperature can be substantially lower than that in thermal CVD. The CNTs have been deposited from various plasma techniques such as hot filament PECVD, [111-114] microwave PECVD, [115-119] d.c. (glow discharge) PECVD, [120-122] and inductively coupled PECVD [123,124] and rf PECVD.[125-127] It is clearly evident from these methods that PECVD is a high yield and controllable method of producing vertically aligned CNTs. The details of the structure of CNTs are outlined in section 4.

### 2.3.3 Catalytic Pyrolysis of Hydrocarbon

This method is commonly used for the bulk/mass production of CNTs by CVD. The main advantage of using this technique is that purification is not required to recover CNTs from the substrate. The simplest method is to inject catalyst nanoparticles (e.g. in the form of a colloidal/particle suspension or organometallic precursors with a carbon feedstock) directly into the CVD chamber. Organometallic compounds (e.g. metallocenes, iron pentacarbonyl and iron (II) phthalocyanine) are often used as precursors for the catalyst.[128-134] These precursors on heating usually get sublimed and catalyst nanoparticles are formed in situ when the compound is decomposed/reduced by heat or hydrogen. A double stage furnace is typically needed because of the different temperatures needed for organometallic sublimation and nanotube growth. In general, the sublimation of metallocenes offers little control over the structural parameters of the nanotubes such as length and diameter. However, it has been shown that by varying the reactive concentration of the metallocene to carbon in the gas phase the average diameter of the structures may be changed. [134,135]. An improvement over the double stage furnace is to use a syringe pump and atomizer to continuously feed a metallocene-liquid carbon (e.g. benzene, xylene, toluene and n-hexane) feedstock solution into a single stage furnace where nanotube growth occurs. [134, 136-139] In almost all cases, the nanotubes are grown on quartz ($SiO_2$), in the form of either a specific substrate or the reactor wall. The advantage of this method is that the aligned CNTs bundles are produced in one step, at a relatively low-cost, without prior preparation of substrates.

The earliest report of such a process-involved pyrolysis of mixture containing benzene and an organometallic precursor e.g. metallocene (such as ferrocene, cobaltocene or nickelocene). [129] In the absence of metallocene, only nanospheres of carbon are found to result. However, a small amount of ferrocene yielded large-quantities of MWNTs. Under controlled conditions of pyrolysis, dilute hydrocarbon-organometallic mixtures yield SWNTs. [130] Gas-phase pyrolysis of $C_2H_2$ along with a metallocene or a binary mixture of metallocenes (Fe, Co, Ni) in flowing Ar or Ar+$H_2$ at 1100$^o$C yields SWNTs. The pyrolysis of a iron pentacarbonyl [$Fe(CO)_5$] - $C_2H_2$ mixture in Ar at 1100$^o$C also produces SWNTs. The diameter of these nanotubes is generally around 1nm, showing that the organometallic precursors give rise to fine particles essential for the formation of such nanotubes.

Recently, CNTs with controllable diameters from ~1 to 200 nm were synthesized by pyrolysis of iron phthalacyanine (FePc) at ~900$^o$C under argon gas flow.[140] The diameters of CNTs have been controlled by varying the metal concentration, using the solid phase dilution of FePc with metal-free phthalocyanine to various degrees. Self-assembled SWNTs have been obtained by using FePc precursor diluted with metal-free phthalocyanine in 1:24 molar ratio. For the production of CNTs the CVD method, a number of carbon-containing feedstock including $CH_4$, $C_2H_2$ and CO have been used as the carbon source, but these gases were used individually in all cases. These carbon-containing gases are the main components in coal gas in other words; coal gas is a mixture of $CH_4$, CO and other gases such as hydrogen and $C_2$-$C_3$ hydrocarbons. The MWNTs have been successfully prepared from coal gas by CVD technique with ferrocene as catalyst. [141]

The addition of trace amount of thiophene ($C_4H_4S$) to liquid hydrocarbon has also been reported to promote the growth of SWNTs, [139,142,143] although higher concentrations of thiophene were reported to revert the growth back to multi-walled in structure (>5wt%). [143] Cheng et al. [143] prepared ropes of SWNTs bundles by the pyrolysis of ferrocene-thiophene benzene mixtures at 1100-1200$^o$C in the presence of hydrogen gas**.** At present, it is possible to generate SWNTs using pyrolysis of various carbon sources in the presence of metals and/or metal alloys. [144,145] However, a novel production method involving the thermolysis of $Fe(CO)_5$ in the presence of CO at elevated pressures (<10 atm) and temperatures (800-1200$^0$C) was reported to be extremely efficient, and nowadays bulk amounts can be produced using this method. This method, called the HiPco (High pressure CO disproportionation) process, was developed by Nikolaev et al.[146] The average diameter of HiPco SWNTs is approximately 1.1 nm. The current production rates approach 450 mg/h (or 10 gm/day) and nanotube typically have no more than 7 mol% of Fe impurities.[147] The standard running conditions are 30 atmosphere of CO pressure and 1050 $^0$ C reaction temperature. Fig. 4 shows a TEM image of the typical product. In the HiPco process nanotube grow in high-temperature flowing CO on catalytic clusters of Fe. Catalyst is formed in situ by thermal decomposition of $Fe(CO)_5$, which is delivered intact within a cold CO flow and then rapidly mixed with hot CO in the reaction zone.

**2.4 Other Synthesis Techniques for Carbon Nanotubes**

The production of CNTs also can be realized by diffusion flame synthesis, [148-157] electrolysis using graphite electrodes immersed in a molten ionic salts, [158-160] ball milling of graphite[161-164] and heat treatment of a polymer. [165,166] Mainly studies in co-flow diffusion flames with the introduction of a catalyst in the form of nano-aerosol or in the form of solid support have been reported. Yuan et al. [148] analyzed flame-based growth of nanotubes on a Ni/Cr support in laminar co-flow methane air diffusion flames. In related work, Yuan and coworkers [149] used a catalytic support in the form of a stainless steel grid to produce CNTs in an ethylene-air diffusion flame. The CNTs have been synthesized in a methane diffusion flame using a Ni-Cr-Fe wire as a substrate. [150] The catalyst particles were Ni and Fe oxides formed on the wire surface inside the flame. When Fe wire was oxidized using nitric acid more nanotubes could be produced. In the flame method, the combustion of a hydrocarbon gas provides both the high temperature and the pyrolyzed hydrocarbon products required for the growth of CNTs. Merchan-Merchan et al. [151] discovered CNTs in $O_2$ enriched counter-flow methane diffusion flames. The metal catalyst dispersed on $TiO_2$ substrate was used by Vander Wal [152] to generate MWNTs in $C_2H_4$ /$O_2$ and $C_2H_2$ /$O_2$ diffusion Co-flow flames. It was demonstrated that the structure of the obtained nanotubes strongly depends on the catalytic particle shape and chemical composition. Recently, Vander Wal and co-workers [153,154] also reported that single-walled CNTs could be grown in premixed co-flow flames by seeding the fuel line with ferrocene and compositions of metal nitrates serving as catalyst precursors for the formation of nantoubes. It was demonstrated that ferrocene and Fe nanoparticles could yield bundles of self-assembled single-walled CNTs with diameters as small as two nanometers. The rich premixed flame synthesis of CNTs using supported catalyst was optimized by selection of optimal flame conditions including fuel composition and fuel-to-air ratio.[155] The studied fuels include methane, ethane, ethylene, acetylene and propane. A recent work by Height and co-workers [156] used premixed co-flow flames to grow SWNTs. Lee et al. [157] prepared CNT on a catalytic metal substrate using an ethylene fueled inverse diffusion flame. The CNTs with diameters of 20-60 nm were formed on the substrate for the case using a stainless steel substrate coated with nickel nitrate. The CNTs were formed in the region of 5-7 mm from the flame center along the radial direction. The gas temperature for this region was ranging from 1027 $^0$C to 527 $^0$C. In the case of a bare stainless steel substrate, the Fe, Ni and Cr particles originally present in the substrate seemed act as the major catalysts. However, when the substrate coated with nickel nitrate was used, the Ni was a major catalyst on the substrate.

The condensed phase synthesis of CNTs was reported by Wen Kuang Hsu and co-workers in 1995. [158] An electric current passing between graphite electrodes, immersed in molten LiCl (electrolyte) at 600$^o$C under an inert atmosphere, resulted in the formation of a mixture of carbonaceous material containing 20-30% yield of CNTs. In this synthesis process it was possible to isolate the CNTs, encapsulated particles and amorphous carbon by filtration, after dissolving the electrolyte in water. Subsequent studies carried out at Cambridge University indicated that the nanotube synthesis strongly depends on the molten salt and the temperature of the electrolyte.[159, 160] In particular, Chen et al.[160] suggested that nanotube formation is caused by graphite intercalation processes occurring at the cathode. Various salts have been successful in the production of nanoutbes by electrolysis: LiCl, KCl, NaCl, LiBr etc. Matveev et al. [167] demonstrated the electrochemical synthesis of CNTs from acetylene solution in liquid ammonia at temperature 233K, which is the lowest ever reported for CNT growth. It is known that liquid ammonia is a good solvent for many organic compounds, e.g. acetylene and nitromethane and at the same time it has a remarkable capacity to stabilize radicals. [168] These features of liquid ammonia provide a unique opportunity to obtain a hydrocarbon radical 'solution', if one chemically or electrochemically initiates chain radical reactions in a solution of hydrocarbons in liquid ammonia. The liquid-phase electrolytic method has not been widely used probably because the nanotubes yield is difficult to control. It is proposed that chain radical reactions may be involved in the CNT growth process. Large amounts of CNTs were synthesized by reduction of $CO_2$ with metallic Li and 550$^{o}$C. [169] In this process, $CO_2$ was used as the carbon source and metallic Li as a reductant to synthesize CNTs. Qian et al. [170] have reported the formation of CNTs by the decomposition of liquefied petroleum gas (LPG) containing sulfur in the presence of $Fe/Mo/Al_2O_3$ catalyst. Prokudina et al.[171] have also reported CNT synthesis from LPG, but no information on sulfur was given.

Ball milling and subsequent annealing is another simple method for the production of CNTs and could be the key to cheap methods of industrial production.[161-164] Preparation of CNTs by ball milling was introduced in 1999 by the Australian group led by Chen[161] and by Huang et al.[163] Chen et al.[162] showed by X-ray diffraction patterns (XRD) that the milling contaminates the graphite powder with Fe from the steel balls. It is this fortuitous Fe contamination that seems to act as the catalyst for the formation of CNTs. Although it is well established that mechanical attrition of this type can lead to fully nanoporous microstructures, it was not until recently that nanotubes of carbon and boron nitride were produced from these powders by thermal annealing.[172] The CNTs have been synthesized by heat-treating the polymer at 400$^{o}$C in air, which was obtained by polyesterification between citric acid and ethylene glycol. [165] The CNTs with the wall thickness from several to more than 100 carbon layers have been prepared by hydrothermal treatment of polyethylene at 800°C in the presence of Ni under 60-100 MPa pressure. [166]

In 2002, Wu et al. [173] have produced CNTs from picric acid. They reported the formation of relatively homogenous (monodisperse) CNTs in very high yields around 80%. Utschig et al. [174] have obtained CNTs by explosive decomposition of an energetic precursor, 2,4,6-triazido-s-triazine ($C_3N_{12}$) in the presence of transition metals Fe, Ni, Cu and Ti. The MWNTs with diameters from 50 nm to 150 nm and verity of morphologies were found. Solutions of transition metal cluster compounds were atomized by electro hydrodynamic means and the resultant aerosol was reacted with ethyne in the gas phase to catalyze the formation of CNTs. [175] The use of an aerosol of iron pentacarbonyl resulted in the formation of MWNTs, mostly 6 nm to 9 nm in diameter, whereas the use of iron pentacarbonyl gave results that were concentration dependent. High concentrations resulted in a wide diameter range (30 nm to 200 nm) whereas lower concentrations gave MWNTs with diameters of 19 nm to 23 nm. The CNTs were synthesized via a novel route using an iron catalyst at the extremely low temperature of 180°C. [176] In this process, carbon suboxide was used as carbon source, which changed to freshly form free carbon clusters through disproportionation. The carbon clusters can grow into nanotubes in the presence of Fe catalyst, which was obtained by the decomposition of iron carbonyl [$Fe_2(CO)_9$] at 250°C under $N_2$ atmosphere.

## 3. PURIFICATION OF CARBON NANOTUBES

As-grown CNTs sample coexists with other carbon species such as amorphous carbons, carbon nanoparticles and transition metals that were introduced as a catalyst during the synthesis.[7,61,69] Several attempts have been tried to purify the CNT powders. Gas phase reaction or thermal annealing in air or $O_2$ atmosphere has been attempted, although the yield of final product (e.g. pure CNTs) was relatively low.[177,178] The key idea with these approaches is a selective oxidative etching process, based on the fact that the etching rate of amorphous carbons and carbon nanoparticles is faster than that of CNTs. Since the edge of the CNTs can be etched away as well as carbonaceous particles during the annealing, it is crucial to have a keen control of annealing temperatures and annealing times to obtain high yield, although the yield is also dependent on the purity of the original sample. Liquid phase reaction in various acids has been tried to remove transition metal.[179-181] This process involves repeated steps of filtering and sonications in acidic solution, where the transition metals were melted into the solution. CNTs are usually cut into small lengths and sometimes broken completely. Therefore, the choice of acids, the immersing time and temperature are the key factors to have high yield, while maintaining the complete wall of CNTs. When using a treatment in $HNO_3$, the acid only has an effect on the metal catalyst. It has no effect on the SWNTs and other carbon particles. [182,183]

Purification of CNTs also was achieved by combining wet grinding, hydrothermal treatment and oxidation processes.[184] The acid reflux oxidation procedure was first described by Rinzler et al., [185] who refluxed raw nanotube materials in nitric acid to oxidize metals and unwanted carbons. They described this as a readily scalable purification process that is capable of handling SWNT material in large batches. Dillon et al. [186] described an oxidation process that produces >98 wt% pure SWNTs. In their purifification process, raw nanotube soot first undergoes a $HNO_3$ fefluxing process.Oxidation of the acid-treated product is then carried out in air at 550$^o$C for 30 min,leaving behind SWNTs that have a weight of 20 wt% of the material. Later, Dillon et al. [187] improved the purification procedure and developed three-step purification method that includes another vacuum annealing at 1500$^o$C to recorder the tubes. Chiang et al. [188] developed a purification method that leads to 99.9% pure SWNTs with respect to metal content. It combines the well-known acid reflux treatment with water reflux and a two-stage gas-phase oxidation process. Reproducible high-yield purification process of CNTs was developed by combining two-step process of thermal annealing in air and acid treatment. [189] This process involves the thermal annealing in air with the powders rotated at temperatures of 470$^{o}$C for 50 min, which burns out the carbonaceous particles and an acid treatment with HCl for 24h, which etches away the catalytic metals. Control of the annealing temperature and rotation of the sample were crucial for high yield.

Bandow et al. [190] reported a procedure for one-step SWNTs purification by microfiltration in an aqueous solution in the presence of a cationic surfactant. Using this procedure, they purified a sample containing 76-90% SWNTs. Micro filtration is based on size or particle separation SWNTs and a small amount of carbon nanoparticles are trapped in a filter. The other nanoparticles (catalyst metal, fullerenes and carbon nanoparticles) are passing through filter. The advantage of this method is that nanocapsules and amorphous carbon are removed simultaneously and nanotubes are not chemically modified. Harutyunyan et al. [191] reported a novel technique to purify SWNTs involving the microwave heating in air followed by HCl acid reatments. The method allows the removal of residual metal to a level <0.2 wt%. Multistep physicochemical methods have been designed to purify nanotubes from the smaller particles. These methods are based on the dispersion of CNTs in polar solvents assisted by surfactant (sodium dodecyl sulfact), followed by ultracentrifugation, microfiltration and size exclusion chromatography.[192] The CNTs are pass over a column with a porous material, through which the CNTs will flow. Size exclusion chromatography is an effective and size separation of CNTs. However this method cannot be scaled up to obtain larger volumes of samples. It is important to note that these methods alter the structural surface of the tubes, which may result in abrupt changes in the electronic transport and mechanical response. Therefore, single-step processes for producing clean nanotubes should be targeted in the future so that chemical treatments and modifications do not alter the fascinating properties of CNTs.

## 4. SYNTHESIS OF ALIGNED CARBON NANOTUBES

Growing organized CNTs on large-scale surfaces is important for obtaining scaled-up functional devices for use as scanning probes and sensors, as field emitters in nanoelectronics and in several other applications. The first group to align MWNTs was lead by Ajayan. [193] They generated aligned CNT arrays by cutting thin slices (50–200 nm thick) of a nanotube-polymer-composite. However, this type of alignment depends on the thickness of the composite slice, and the process is impractical for generating larger areas of aligned tubes. Another possibility of aligning MWNTs (produced in the arc-discharge) is by direct rubbing of the nanotube powder on a substrate. Another approach to aligning the CNTs has been reported by de Heer et al.,[194] who prepared a uniform black deposit by drawing a nanotube suspension through 0.2μm pore ceramic filter and then transferring the deposited material onto a plastic surface (e.g. teflon) to form a film of aligned CNTs.

Aligned CNTs have been obtained by the CVD of $C_2H_2$ catalyzed by Fe nanoparticles embedded in mesoporous silica, the pores of the silica controlling the growth direction of the CNTs. [195,196] Terrones et al. [197] have reported the formation of aligned CNTs by the pyrolysis of 2-amino-4, 6-Chloro-s-triazine over thin films of a Co catalyst patterned on a silica substrate by laser etching. This method offers control over length (up to 50 μm) and fairly uniform diameters (30-50 nm), as well as producing nanotubes in high yield and uncontaminated by polyhedral particles. Nath et al. [198] have prepared aligned CNT bundles by the pyrolysis of $C_2H_2$ over Fe or Co nanoparticles well dispersed on silica substrates. The catalyst was silica-supported Fe-Co prepared by sol-gel method. Fan et al.[199] also synthesized "towers" of densely packed nanotubes by using a 5 nm thin film of evaporated Fe on electrochemically etched porous silicon. The key factor in achieving dense, aligned growth of CNTs is the preparation of dense and active catalyst particles on the substrate surface. Kong et al. [200] showed catalyst islands fabricated on Si wafers using electron beam lithography could be used to grow (by CVD) SWNTs oriented in the plane of the substrate. Aligned CNT arrays was grown on the $Fe_2O_3$ /$SiO_2$ / Si substrates by the pyrolysis of ethanol as carbon source at 700°C. [201] Aligned CNTs arrays were synthesized in the presence of Co nanoparticles. [202] By taking advantage of the fact that Co nanoparticles are dispersible in organic solvent, an ink-jet technique has been applied to the Co nanoparticle dispersion for CNT patterning.

Another common means of achieving aligned growth is through the use of templates, the most popular of which are vertical nanopores created by the electrochemical processing (anodization) of aluminum. Porous alumina membranes typically contain vertical nanopores, which are a few to hundreds of nanometers in diameter and lengths, which can range from a few micrometers to hundreds of micrometers. The use of porous alumina in the synthesis of CNTs was first reported in 1995. [203] In general, two types of template grown structures are possible, namely catalyzed and pyrolytic (no catalyst). The latter usually requires higher temperatures in order to decompose the carbon feedstock gas. Che et al. [204] prepared CNTs with diameters ~20 nm using an alumina template in ethylene/pyrene with a Ni catalyst at 545$^{o}$C or without catalyst at 900$^{o}$C. After growth, the alumina template can be removed by dipping in HF or NaOH solution to reveal an array of well-ordered CNTs standing perpendicular on the substrate. Li et al.[205] prepared aligned CNTs using an alumina template together with Co/Ni catalysis of $C_2H_2$ at 650$^{o}$C. The smallest nanotubes ~0.4 nm in diameter, have been fabricated using zeolite templates. [206]

Another approach that has been successfully demonstrated is the use of plasma-assisted hot filament chemical deposition in conjunction with plasma-assisted modification of the catalyst surface (such as Ni deposited on Si). [113]. Large-scale arrays of vertically aligned (to the substrate) CNTs have been fabricated by this method. Aligned MWNTs were grown normal to Ni-coated glass substrates below 666$^{o}$C by plasma-enhanced hot filament CVD of a mixture of $C_2H_2$ and $NH_3$. [111] Huang et al. [207] studied thin film Co, Fe and Ni catalysts on a Ti substrate using the hot filament dc-PECVD process and found that the CNTs produced with Ni were structurally the best in terms of graphitization, straightness, lack of amorphous carbon over coating and structural defects such as openings in the walls. The CNTs nucleated from Ni also had the fastest growth rate. They noted that the diameters of the Ni catalyzed CNTs were larger than the CNTs catalyzed by Co and Fe, indicating that Ni had the weakest interaction with the Ti substrate and hence formed the largest catalytic clusters. Recently, large periodic arrays of CNTs have been grown by plasma enhanced hot filament CVD on periodic arrays of Ni dots that were prepared by polystyrene nanosphere lithography. [208] A single layer of self-assembled polystyrene spheres was first uniformly deposited on Si wafer as a mask, and then electron beam vaporization was used to deposit a Ni layer through the mask. A microwave PECVD system was used to grow CNTs catalyzed from a Co thin film on a Si substrate. [115] The CNTs were deposited using PECVD and then the plasma was stopped for conventional thermal CVD to continue. The resultant nanotubes were thus straight for the plasma-grown section and curly for the thermally grown section. Additionally, when non-planar or angled substrates were introduced into the plasma, the CNTs still grew perpendicularly from the substrate surfaces because the microwave plasma formed a sheath around these objects. Tsai et al., [209] who used microwave plasma to synthesize, aligned CNTs, proposed as model based on anisotropic etching in order to explain the vertical alignment of the CNTs. They suggested that nanotubes, which grew in random orientations, were unprotected by their metal catalyst particle and were hence anisotropically etched away in the plasma.

Chhowalla et al. [121] performed a parametric study of the dc-PECVD growth of CNTs using $C_2H_2$ and $NH_3$ gases with Ni catalyst. They studied the effect of the initial catalyst thickness, the effect of the $C_2H_2$ ratio in the gas flow, the effect of pressure, deposition time and the effect of the dc bias on the substrate. By optimizing the plasma characteristics, it was determined that 0.15V/μm was the minimum electric field in the plasma sheath necessary for vertical alignment of the CNTs. Clearly, to obtain a high degree of vertical alignment, one should thus maximize the electric field in the plasma sheath by increasing the substrate bias or by increasing the gas pressure (which increases ionization, leading to a higher field in the sheath). Teo et al. [210,122] used the dc PECVD process to demonstrate the high yield and uniform growth of patterned areas of CNTs and individual CNTs. In patterned growth especially, the $C_2H_2$ to $NH_3$ ratio was found to be important in achieving amorphous carbon free growth. During growth, the $C_2H_2$ is continuously being decomposed by the plasma (at a rate much faster than thermal pyrolysis) to form amorphous carbon on the substrate surface and role of $NH_3$ in the plasma is to etch away this unwanted amorphous carbon. The N and H species in the $NH_3$ plasma react with the amorphous carbon to form volatile C-N and C-H species. Hence there exists an optimum condition, which was determined to be a flow ratio of 40:200 sccm of $C_2H_2$: $NH_3$ at 700°C, where the production and etching of amorphous carbon is balanced, thus yielding substrates which are free of amorphous carbon.

Regular arrays of freestanding single CNTs were prepared on Ni dot arrays by dc plasma-enhanced CVD.[211] The vertical alignment of a single CNT was directly dependent on the location of the catalyst metals. Dielzeit et al. [124] performed a parametric study of vertically aligned CNT growth in inductively coupled plasma. They used thin film Fe catalyst that was deposited on an aluminum layer on Si substrates and the CNTs were grown using $CH_4$, $H_2$ and Ar gases. Aligned CNTs have been synthesized by hot-filament CVD on Ni film coated Si substrate. [212]. These CNTs were well-aligned perpendicular to the substrate. In this technique, the filament, which is typically placed near the substrate, acts as a thermal source for decomposing the gas, and often serves to heat the substrate as well.

Another strategy of obtaining aligned CNTs growth is through the use of electric fields. Avigal et al. [213] used a vertical electric field during growth to achieve vertical alignment of CNTs. Aligned CNTs were grown by electric arc discharge of graphite rods under helium atmosphere and applying the electric field during the growth process. [214] The electric fields were varied as 3, 6, 9, 12, 15 and 21 volts during the growth process of CNTs in the carbon plasma. The best results have been obtained with electric field corresponding to 6V where the as-formed tubes are in parallel alignment and exist as bundles. As the electric field is increased, the alignment of tubes in the bundle becomes randomly oriented. The degree of randomness increases with increase of electric field after its optimum value corresponding to 6V. A possible factor influencing the growth of parallel CNTs will be the alignment of tube axis during its growth along the direction of the applied electrical field. These directions are along the length of the graphite electrodes perpendicular to their diameter. The experimental conditions prevalent involving application of external field while the electrode tip was very hot, is suggestive of the fact that the applied field leads to alignment of tubes during growth leading to CNT bundles. Applying the standard model as developed by Rasch [215] for evaluation of electric field under plasma arc in a gaseous ambient, it has been found that the field for 6 volts corresponds to 1888.5 volts/cm. [216] Lateral electric fields can also be used to guide nanotubes during growth. Zhang et al. [217] prepared electrode and catalyst “fingers” on a quartz substrate which were biased during CVD in order to create a lateral electric field. They found that electric fields of 0.13-0.5 volt/μm were needed to guide and align SWNTs during growth. Lee et al. [218] also reported a technique for the directed growth of lateral CNTs. They prepared a sandwich structure comprised of $SiO_2$-Ni-Nb on Si. Using microfabrication, only one face of the sandwich was left exposed to the gases and it was from this face that the Ni catalyzed the outward/lateral growth of CNTs in $C_2H_2/N_2$ at 650°C. Zhang et al. [219] have reported the formation of aligned CNTs by simple electrically heating a substrate in organic liquids. Si substrates with a thin Fe film cover were electrically heated to 500-1000°C in organic liquids. Aligned CNTs have been grown and have been found to adhere well on the Si substrates by using methanol and ethanol. The top ends of the CNTs were closed with seamless caps. Some other shaped CNTs have also been synthesized, in particular, a special kind of coupled CNT of which the structure of one CNT is left hand rotated and the other right hand rotated with bridging chains between them.

Self-assembly pyrolytic routes, which avoid patterning of any substrate, have also been developed in order to obtain large and continuous arrays of MWNTs. Wang et al. [220] and Araki et al. [221] grew pillar-like structures consisting of aligned CNTs by heating powders of iron (II) phthalocyanine at $550^0$C under an Ar-$H_2$ atmosphere. Nerushev et al.[222] successfully demonstrated that the simple thermolysis $Fe(CO)_5$ resulted in the production of uniform films of aligned MWNTs. The thermolysis of solid carbon-containing precursors (naphthalene, anthracene, pyridine, etc.) in the presence ferrocene (or mixtures with other metallocenes) yields aligned CNTs .[223] Good quantities of aligned CNTs bundles have been prepared by the gas phase (vapor) and liquid phase pyrolysis of metallocene-hydrocarbon mixture.[134,137,139,224-227] Aligned CNTs bundles have been prepared by the gas (vapor) phase pyrolysis of a mixture of ferrocene and $C_2H_2$ at $1097^o$C in flowing Ar. [224] The prevalent method is using chemical dissociation of ferorcene (containing carbon and catalyst Fe) in conjunction with some hydrocarbon compound e.g. $C_2H_2$, $C_2H_4$, benzene, xylene etc. The resulting CNTs invariable carry Fe embedded in CNTs. For application purposes, for example, for electronic application or storage material, pure CNTs devoid of catalytic particle (e.g. Fe) filling are required. The CNTs without Fe inclusion and in aligned configurations have been prepared by the gas phase pyrolysis of ferrocene in the presence of $C_2H_4$. [225] This has been achieved through optimization of growth parameters, for example, heating rate of ferrocene, pyrolysis temperature, flow rates of Ar and $C_2H_4$. Table 2 gives an overview of the significant differences regarding the morphology of the CNTs. The optimum results relating to synthesis of CNTs without Fe inclusion and in aligned configurations were obtained at $1000^o$C pyrolysis temperature under flow rates of Ar ~1000 sccm (standard cubic cm per min). The cause for the alignment of CNTs is a dense orderly distribution for the Fe metal atoms. This allows a dense aligned nucleation of CNTs, leading to the formation of aligned tubes, which aggregate in bundles.

In liquid phase pyrolysis, solution containing both the hydrocarbon source (xylene, benzene etc.) and metal catalyst precursor (ferorcene) were injected into the furnace with a syringe and subsequently pyrolyzed, thus producing bulk amount of aligned CNTs. [134,137,139,226,227] Kamalakaran et al. [137] prepared thick CNT arrays by pyrolysing a jet (spray) solution of ferrocene and benzene in an Ar atmosphere at 850$^o$C temperature.Fig.5 (a) illustrates large carpet-like areas of aligned nanotubes. Fig. 5 (b) depicts arrays consisting of thick tubes exhibiting diameters <200 nm. Tube diameter, degree of alignment and crystallinity can be controlled by varying the Ar flow rate and the Fe:C ratio within the precursor solution. This low cost route for the synthesis of CNTs is advantageous due to the absence of $H_2$ as a carrier gas and the low pyrolytic temperature. Aligned CNTs were also prepared by the spray pyrolysis of ferrocene benzene solution. [226] In Table 3, we summarize the nature of products obtained by the spray pyrolysis of ferrocene benzene solution. Fig. 6(a) reveals that the CNTs exist in the form of bundles (≤0.01mm$^2$) made up of aligned CNTs These bundles made up of aligned CNTs. These bundles made up of CNTs are found to exist in specific local regions often separated from each other forming soot like configuration and without definite shape. The average length of CNTs is 100 µm. The optimum growth parameter is given in Table 3 (S.N. 1). The magnified view of one of the bundle is shown in Fig.6 (b). It shows the formation of dense and pure aligned CNT having outer diameter ranging from ~ 60 nm to 100 nm.

High purity aligned CNTs films were grown on quartz substrates by injecting a solution of ferrocene in toluene into a suitable reaction furnace. [134] In this case, by varying the growth temperature, CNTs of various packing densities were produced. Aligned CNTs grows within the temperature range 590 to 850$^o$C, with a maximum yield at 760$^o$C. The diameter and diameter distribution of the nanotube increased with increasing temperature and ferrocene concentration. The number of graphitic defects present within the nanotubes was shown to be dependent on temperature and the concentration of ferrocene within the toluene solution. The concentration of encapsulated material within the nanotubes increases as the temperature and the ferrocene increases and at the extremes of time and temperature explored, an overgrowth of fine particles and fibrous materials appears on top of the aligned nanotube films. Wei and coworkers [227] noticed that by pyrolyzing solutions of ferrocene and xylene-aligned CNTs grow preferentially perpendicular to the $SiO_x$ substrates. Subsequently, the same group used photolithographic techniques to pattern Si substrates with multiple $SiO_2$ structures and, for the first time, were able to grow multidirectional aligned nanotube arrays (e.g., pillars, cylindrical ropes, square ropes, flower-like structures, etc.) (Fig.7). Zhu et al. [139] fabricated extremely long bundles of SWNTs of several centimeters (<20 cm), using n -hexane as carbon source in conjunction with ferrocene and thiophene in the presence of $H_2$ at 1150 $^0$C.

Mayne et al. [138] developed a sophisticated aerosol generator in order to pyrolyze homogeneously dispersed aerosols generated from benzene/ferrocene solutions, at 800 - 950 $^0$ C. This method, involving the pyrolysis of aerosols, opens up new avenues in CNT synthesis using liquid hydrocarbons and catalyst precursors. The advantage of this process is the continuous and simultaneous feeding of the reactor with homogeneous hydrocarbon/catalyst aerosols, which results in high yields of aligned CNTs. Subsequently, Grobert et al. [228] showed that the thermolysis of organometallic mixtures (e.g., ferrocene-nickelocene) in benzene solutions, using the aerosol generator, yields extremely clean flakes consisting of aligned CNTs. Recently, CNTs with fairly uniform diameter and aligned CNT bundles have been prepared by nebulized spray pyrolysis using ferrocene-hydrocarbon (acetylene, benzene, toluene, xylene, n-hexane) mixture. [229] Nebulized spray is a spray generated by an ultrasonic atomizer. The quality of the product is dependent on the pyrolysis temperature, carrier gas flow rate, additional carbon sources used and the catalyst precursor concentration. Many of the hydrocarbons yield aligned MWNT bundles and SWNTs were obtained in certain instances. Well-graphitized MWNTs were obtained when xylene was used as the additional carbon source. Nebulized spray pyrolysis of Fe $(CO)_5$ in the presence of $C_2H_2$ also yields aligned CNT bundles.

## 5. SYNTHESIS OF SPECIAL CARBON NANOTUBES CONFIGURATIONS: NANOCOILS, NANOHORNS, BAMBOO-SHAPED CNTs AND CARBON CYLINDER MADE UP FROM CNT

Carbon structures have been a subject of extensive research since the synthesis of the form of carbon: fullerenes and CNTs. [1,3] In recent years, considerable efforts have been made to fabricate different CNT morphologies and explore their application. Carbon nanotubes with helical /spiral form have received much attention recently because of their unique helical/spiral structures and their special mechanical and electromagnetic properties. Many efforts have been attracted in the synthesis and the property study of carbon coils of micrometer scale size, which are also called carbon micro-coils. [230-233] With the development of nanotechnologies, carbon coils of nanometer scale size or carbon nanocoils are expected in the fabrication of nanodevices such as a generator or detector of magnetic field, an inductive circuit, an actuator, a spring etc. Hernadi et al. [234] reported on the formation of nanotubular coils by Fe catalyzed pyrolysis of $C_2H_2$. The carbon coils have been prepared by the Ni-catalyzed pyrolysis of $C_2H_2$ containing a small amount of thiophene as an impurity.[232] Wen et al.[235] prepared coiled CNTs by Ni-catalyze pyrolysis of $C_2H_2$. The coiling can result from the introduction of pentagon-heptagon pairs at regular distances in the hexagonal network forming the wall of a straight CNT.[236] Carbon nanocoils have been synthesized at high yield by catalytic thermal decomposition of $C_2H_2$ gas using Fe and indium tin oxide (ITO) as the catalyst. [237] The coils generally consist of two or more nanotubes. Each coil has its own external diameter and pitch, which is determined by the structure of the catalyst at its tip. The effect of Fe and ITO on the growth of carbon nanocoils has been investigated. [238] By increasing the content ratio of tin in the ITO film, the yield of CNTs may be increased but that of nanocoils is decreased. It is found that Fe-additions lead to the growth of CNTs, while ITO induces their helical growth.

Carbon nanocoils with twisting form have been synthesized by the $Ni/Al_2O_3$ catalyzed pyrolysis of $C_2H_2$. [239] Recently, single-helix twisted carbon nanocoils and single-helix spring-like carbon micro/nanocoils have been prepared by the CVD process of the catalytic pyrolysis of $C_2H_2$ at 700-800°C over Fe-containing alloys in large scale.[240] The results indicated that most of the twisted nanocoils grown by a two directional growth mode; that is two twisting nanocoils grew out of a catalyst grain in opposite chirality. Spring like carbon coils have been prepared with over sputtered Fe alloys on ceramics supporters, with a high purity and good reproducibility.[240] All of the spring like carbon coils is of one directional growth mode and their coiling diameter and coil pitch are about the same size of several hundred nanometers.

Single-wall carbon nanohorns (SWNHs) are a new type of carbon material having a horn-shaped sheath of single-wall graphitic sheets and they associate with each other to form a 'Dahlia-flower' like aggregate.[241] Single-walled carbon nanotubes and nanohorns have been fabricated by means of a torch arc method in open air.[242] A graphite target containing Ni/Y catalyst was used as a counter-electrode of the welding arc torch. The target was blasted away by the d.c. arc and soot was deposited on the substrate placed downstream of the arc plasma jet. Wang et al. [243] prepared SWNHs by arc discharge of graphite electrodes submerged in liquid nitrogen. The product in its powder form was found to consist of spherical aggregates with sizes in the range of 50-100nm. A novel nanocarbon structure, carbon nanohorns particle that includes one Ni-contained carbon nanocapsule in its center, is produced by submerged arc in liquid nitrogen using Ni-contained (0.7 mol%) graphite anode.[244] A method of synthesizing bulk amounts of SWNHs on a significant low-cost basis is proposed based on the arc in water method with $N_2$ injection in to the arc plasma.[245] It is elucidated that rapid quenching of the carbon vapour in an inert gas environment is necessary to form the delicate SWNH structure. To realize this reaction field, the arc plasma between the graphite electrodes was isolated from the surrounding water by a thin graphite wall with and $N_2$ flow that excluded reactive gas species ($H_2O$, CO and $H_2$) from the arc zone. High concentrations of SWNHs were found as fine powders floating on the water surface.

A multi-walled CNT with a bamboo-like structure is one of the familiar nanotubes. The bamboo-like MWNT can be prepared by various methods such as d.c. arc discharge, thermal CVD and microplasma assisted CVD.[246-253] Lee et al. [248] have reported the growth of vertically aligned CNTs on Fe-deposited $SiO_2$ substrate by thermal CVD of $C_2H_2$ at 750-950$^0$C. All the CNTs have no encapsulated catalytic particles at the closed tip and a bamboo structure. Ma et al.[251, 252] suggested that the bamboo-like structure is formed by creating a positive curvature in the closed nanobell structure when a five-membered ring and the catalyst metal nanoparticle serves only as an initial nucleation center. However, Suenaga et al.[254] measured the nitrogen intensity profile across the tube axis of a MWNT with truncated cone-shaped graphitic shells and argued that the accommodation of nitrogen in the graphitic intrashells cannot be confirmed. Bamboo-like CNTs have been grown on silicon substrates by microwave plasma enhanced CVD using $C_2H_2/NH_3$ mixtures.[253] The majority phase is found to have a bamboo-like structure with concentric hollow structures also in existence. Bamboo-like CNTs have been prepared by pyrolysis of Ni-phthalocyanine.[255]

Over the past decade of nanotube research, a variety of organized nanotube architectures have been fabricated using CVD technique.[111,138,195,227,256,257] The idea of using nanotube structures in separation technology has been proposed, but building macroscopic structures that have controlled geometric shapes, density and dimensions for specific applications still remains a challenge.[258-260] In liquid phase pyrolysis experiment soot containing CNTs were coated on the surface of quartz tube forming a thin film (mat like), which motivated to grow bulk objects made up of CNTs.[134,137,226] A micro-object of a specific shape (hollow-cylinder) containing high purity aligned CNTs were prepared by the spray pyrolysis of ferrocene benzene solution.[226] The realization of the same was possible by controlled flow rate of ferrocene benzene solution, reaction time and optimum nozzle dimension of the standard, spray pyrolysis set up. The optimum parameter is listed in Table 3 (S.N. 2). The bulk tube (hollow cylinder) so formed was having the outer diameter identical to the inner diameter of the silica tube and the length was found to be several centimeters (~ 3-4cm). Fig. 8(a) shows the optical photograph of the bulk tube formed by this route. The SEM investigation along the thickness/diameter of the bulk tube reveals the growth of aligned CNT bundles in radial direction [Fig. 8(b)].[31] The growth of the CNTs takes place all along the thickness of the bulk tube. This reveals a highly organized self-aligned growth of CNTs. The magnified SEM studies show the high-density growth of radially aligned CNT (Fig. 9). The average length of CNTs was ~300μm. The length of CNTs corresponds to the wall thickness of the bulk tube.

## 6. SUMMARY

Carbon nanotubes clearly constitute a fascinating class of materials exhibiting a variety of novel properties. Synthesis is the key issue of CNT research. There has been wide variety of techniques for synthesizing CNTs. In this article, the different synthesis methods of CNTs have been reviewed. The CNTs (single and multi-walled) are produced by three main techniques, arc discharge, laser vaporization and chemical vapor deposition. In addition to these three main synthesis routes for CNTs, other methods e.g. diffusion flame, electrolysis of graphite electrodes in molten ionic salts, ball-milling of graphite, heat treatment of polymer etc have also been used. The CNTs from the arc discharge are often covered with amorphous carbon, which contains metallic particles in the case of metal-carbon co-evaporation. The yield of CNTs from arc discharge is not very high, whereas with laser vaporization, the yield is much higher, but the quantities are small. With the laser vaporization method, CNTs are very clean i.e. less covered with amorphous carbon. The yield and shape of CNTs prepared by laser vaporization are determined by fewer parameters than are tubes obtained by arc discharge synthesis. Catalytic CVD is an extremely versatile technique for the production of CNTs. One major advantage of the CVD approach is that CNTs can be made continuously, which could provide a very good way to synthesize large quantities of CNTs under relatively controlled conditions. A variety of hydrocarbons, catalyst and catalyst supports have been used successfully by various groups worldwide to synthesize CNTs. Many investigations into the properties of CNTs and their potential applications require clean and uniform CNTs that contain no impurities. A number of purification methods have been developed to date. Care should be taken when the technique is chosen, as the effect on the entire sample will also depend on the composition and the amount of the sample. In last few years, significant progress has been made for fabricating various alignments and patterns of CNTs. The CVD technique can also be adopted for the controlled growth of CNTs at particular sites on a substrate for various applications. Due to the great potential of CNTs, it is clear that novel technologies will emerge in the near future.

**Acknowledgements**

The authors are extremely grateful to Prof. A.R. Verma, Prof. C.N.R. Rao, Prof. P. Ramachandra Rao, Prof. S. Lele, Prof. P.M. Ajayan, Prof. H.S. Nalwa and Prof. T.V. Ramakrishnan for their encouragements and support. Helpful discussion with Prof. R.S. Tiwari is gratefully acknowledged. The authors acknowledge with gratitude the financial support from MNES, UGC, CSIR (New Delhi) and DST(NSTI).

Table 1. Summary of Synthesis of CNTs using CVD Techniques

| Catalyst and substrate | Carbon source | Reaction Condition | | Product Quality | Ref. |
|---|---|---|---|---|---|
| | | Temp. ($^0$C) | Flow rate of gases (ml/min) | | |
| Fe, Co/Y, $SiO_2$ | $C_2H_2$ | 700 | | MWNTs (76%) | 80 |
| Co, Fe, Cu/$SiO_2$, Zeolite, clay | $C_2H_2$ | 700 | $C_2H_2$ ~8 $N_2$ ~75 | MWNTs (71-76%) and SWNTs (low yield) | 81 |
| Co-Mo, Co-V, Co-Fe/$SiO_2$, Zeolite | $C_2H_2$ | 700 | $C_2H_2$ ~30 $N_2$ ~300 | MWNTs | 82 |
| Mo, Quartz | CO | 1200 | CO ~1200 (flow in chamber) The CO pressure was held at ~100 torr for 1 hr | SWNTs (1-5nm in diameter) | 84 |
| Co-Mo (1:2)/$SiO_2$ | CO | 700 | CO ~100 | MWNTs (4%) & SWNTs (88%) | 86 |
| Co, Ni, Fe/MgO | $H_2/CH_4$ (4:1) | 1000 | $H_2$ ~300 $CH_4$ ~75 | SWNTs (70-80%) | 87 |
| Fe-Mo, Fe-Ru/$Al_2O_3$-$SiO_2$ | $CH_4$ | 900 | $CH_4$ ~6000 | SWNTs (~42 wt%) | 95 |
| $Mg_{0.9}Co_{0.1}O$ | $H_2/CH_4$ | 1000 | 18 mol% $CH_4$ $CH_4$ ~250 | SWNTs/DWNTs ~ 64wt% (>80% 0.5 – 5 nm diameter) | 99 |
| Fe/$Al_2O_3$ | $CH_4$ | 900 | - | DWNTs (1-6 nm diameter) | 100 |
| Fe-Mo, Quartz | $CH_4$ | 875 | $CH_4$ : Ar (1:1 ratio) | DWNTs (in high yields) | 101 |
| Single or multi-Fe, Co, Ni/$Al_2O_3$ $SiO_2$ | $C_2H_4$ | 1080 | $C_2H_4$ ~30 $N_2$ ~80 | Bundles of isolated SWNTs (0.7 or 0.2 nm diameter) | 104 |
| Fe, Si-substrate | $C_2H_5OH$ | 900 | Ar+$H_2$ (6%) ~20 | SWNTs (4 cm length) | 105 |
| Fe/Mo, $SiO_2$/Si | $CH_4$ | 900 | $CH_4$ (~1500) $H_2$(~125) | SWNTs | 108 |
| Al/Fe/Mo, Si-substrate | $C_2H_2$ | 1000 | $C_2H_2$ ~50-250 | SWNTs (~1.3nm diameter) | 109 |
| Fe,Co, Zeolite powder<br>Mo,Co Si or quartz substrate | $C_2H_5OH$<br>$C_2H_5OH$ | 800<br>800 | $C_2H_5OH$ vapour(5 Torr),Ar<br>$C_2H_5OH$ vapour(10 Torr), Ar-$H_2$(3%) | SWNTs( >40 wt%)<br>Mat of SWNTs | 110 |
| Metallocene $M(C_5H_5)_2$ (M=Fe, Co, Ni) | $C_6H_6$ | 900 | Ar (75%) $H_2$ (25%) Ar+$H_2$ ~50 & 1000 | MWNTs | 129 |
| $M(C_5H_5)_2$ (M=Fe,Co,Ni), $Fe(CO)_5$ | $C_2H_2$ | 1100 | Ar ~975 $H_2$ ~25 $C_2H_2$ ~50 | SWNTs (diameter ~1 nm) | 130 |
| $(FePc)_{0.25}(H_2Pc)_{0.75}$<br>$(FePc)_{0.08}(H_2Pc)_{0.92}$<br>$(FePc)_{0.04}(H_2Pc)_{0.96}$ | | 900-<br>850 | Ar ~40<br>$H_2$ ~40<br>Ar ~40<br>$H_2$ ~40 | CNTs (high yield)<br>SWNTs (~2.4nm diameter) | 140 |
| Fe $(C_5H_5)_2$, Quartz | Coal gas | 850-950 | Coal ~50-200 | SWNTs (1-2nm diameter) | 141 |
| Fe $(C_5H_5)_2$ | $C_6H_6$ & $C_4H_4S$ | 1100-1200 | $H_2$ ~70-90 & 150-225 | SWNTs (0.5 ~5wt% thiophene addition) MWNTs (higher than 5 wt%) | 143 |
| Fe $(CO)_5$ | CO | 800-1200 | CO (<10 atm) | SWNTs (in high yields) | 146 |
| Fe $(CO)_5$ | CO | 1050 | CO(30 atm) | SWNTs (450 mg/h) | 147 |

**Table 2:** Summary of experiments for ferrocene-ethylene pyrolysis leading to alignment of CNTs without Fe inclusion

| **SN** | **Ferro-cene amo-unt (mg)** | **Pyro-lysis temp. ($^oC$)** | **Ferro-cene heating rate ($^oC$/min)** | **Flow rate of Ar (sccm)** | **Flow rate of $C_2H_4$ (sccm)** | **Anneali-ng time (min)** | **Result** |
|---|---|---|---|---|---|---|---|
| 1. | 300 | 1000 | 80 | 300 | 100 | 10 | CNTs without Fe inclusion (in high yield); Alignment: Random; O.D.: ~25-50 nm |
| 2. | 300 | 1000 | 80 | 500 | 100 | 10 | CNTs with negligible Fe inclusion; Alignment: Random; O.D.: ~20-60 nm |
| 3. | 300 | 1000 | 80 | 1000 | 100 | 10 | CNTs with negligible Fe inclusion; Alignment: Aligned; O.D.: ~20-60 nm |

**Table 3**: Optimum conditions for representative spray deposition Experiments.

| Respective Samples | Concentra-tion of Ferrocene in Benzene (mg/ml) | Nozzle diameter (id-mm) | Diameter of outer orifice (mm) | Temp. °C | Reaction time (min) | Flow rate (ml/ min) | Results |
|---|---|---|---|---|---|---|---|
| S1 | 75 | 0.35 | 0.50 | 850 | 20 | 2 | Soot containing aligned carbon nanotubes bundles |
| S2 | 75 | 0.50 | 1.5 | 900 | 20 | 3 | Bulk tubes comprising of carbon nanotubes |

**Figure Captions:**

Fig. 1: Different forms, (allotropes) of carbon.

Fig. 2: A TEM image of SWNTs synthesized by alcohol catalytic CVD method. Note the absence of metallic impurities and MWNTs. Reprinted from Ref. 110 S. Maruyama et al., Jr. of Nanosc. and Nanotech. 4, 360, (2004) with permission from American Scientific Publishers.

Fig. 3: SEM image of SWNTs directly synthesized on a quartz substrate taken at a tilted angle, including a broken cross-section of the substrate. The generated mat of SWNTs on the quartz surface has a thickness of a few hundred nanometers. Reprinted from Ref. 110 S. Maruyama et al., Jr. of Nanosc. and Nanotech. 4, 360, (2004) with permission from American Scientific Publishers.

Fig. 4: TEM image of the typical SWNTs produced in the HiPco reactor. Reprinted from Ref. 147 P. Nikolaev, Jr. of Nanosc. and Nanotech. 4, 307, (2004) with permission from American Scientific Publishers.

Fig. 5: SEM images of (a) large mat of aligned CNTs formed by the spray pyrolysis of ferrocene under optimized conditions and (b) higher magnification of (a) exhibiting the presence of thick tubes (~150-200 nm) within the arrays. Reprinted from Ref. 137 R. Kamalakaran et al., Appl. Phys. Lett. 77, 3385, (2000) with permission from American Institute of Physics.

Fig. 6: (a) SEM micrograph showing bundle containing aligned CNTs, (b) magnified SEM micrograph of one of the above bundle showing aligned CNTs. Reprinted from Ref. 226 (unpublished).

Fig. 7(a-d) SEM images of CNT arrays grown using pyrolytic processes on silica substrates. In, (a) growth in the vertical direction occurs from the top silica surfaces, (seen as arrays emanating from the center of each pattern), growth on the sides occurs as horizontal arrays, (sideways growth seen an each pattern; the diameter of an individual flower is ~50μm). Reprinted from Ref. 227 and courtesy of B.Q. Wei and P.M. Ajayan et al., Nature 416, 495 (2002) with permission from Nature Publishing Group.

Fig. 8: (a) Optical photograph showing the bulk tube (hollow cylinder) synthesized by spray pyrolysis technique and (b) SEM image of the aligned CNTs with radial symmetry resulting in hollow cylindrical structure (scale 1mm). Reprinted from Ref. 31 A. Srivastava et al., Nature Material 3, 610, (2004) with permission from Nature Publishing Group.

Fig. 9: SEM micrograph showing highly aligned dense CNTs. Magnification: 3000x. Reprinted from Ref 226 (unpublished).

| | | |
|---|---|---|
| Diamond<br>(a) | 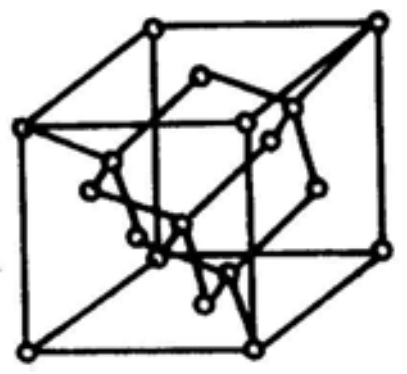 | Cubic (f.c.c.)<br>a = 3.566Å |
| Graphite<br>(b) | 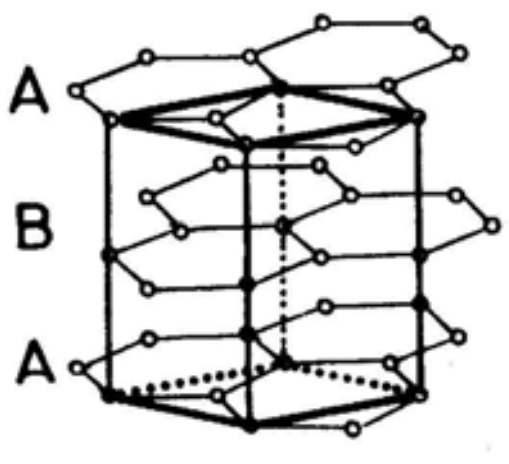 | Hexagonal<br>a = 2.463Å<br>c = 6.714 Å |
| Fullerene<br>(c) | 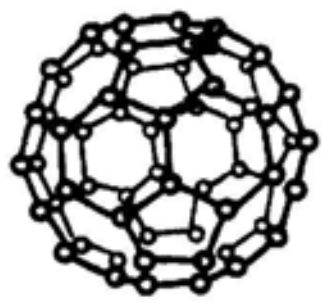<br>**$C_{60}$**<br>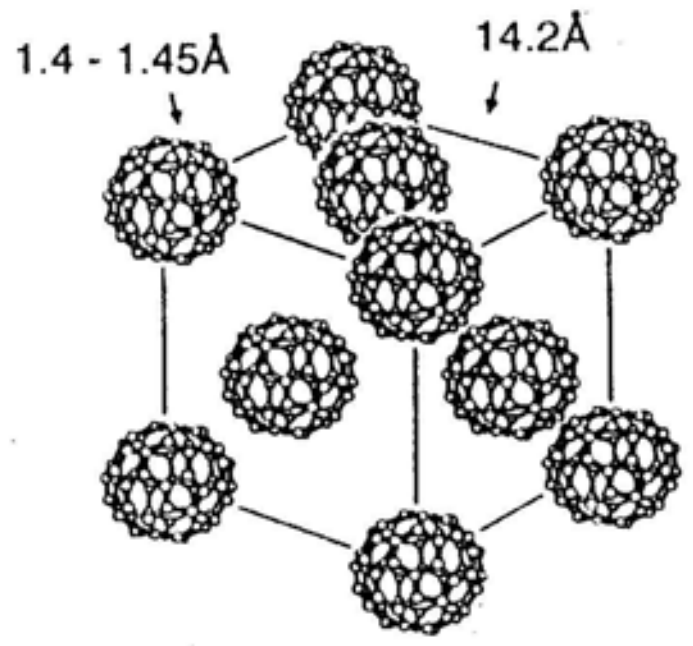 | Cubic (f.c.c.)<br>a = 14.17 Å |
| Carbon Nanotubes (CNTs)<br>(d)<br>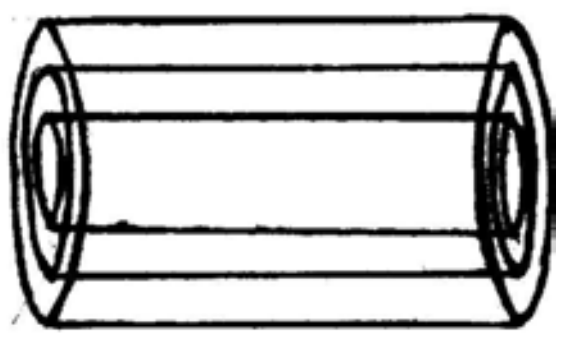 | 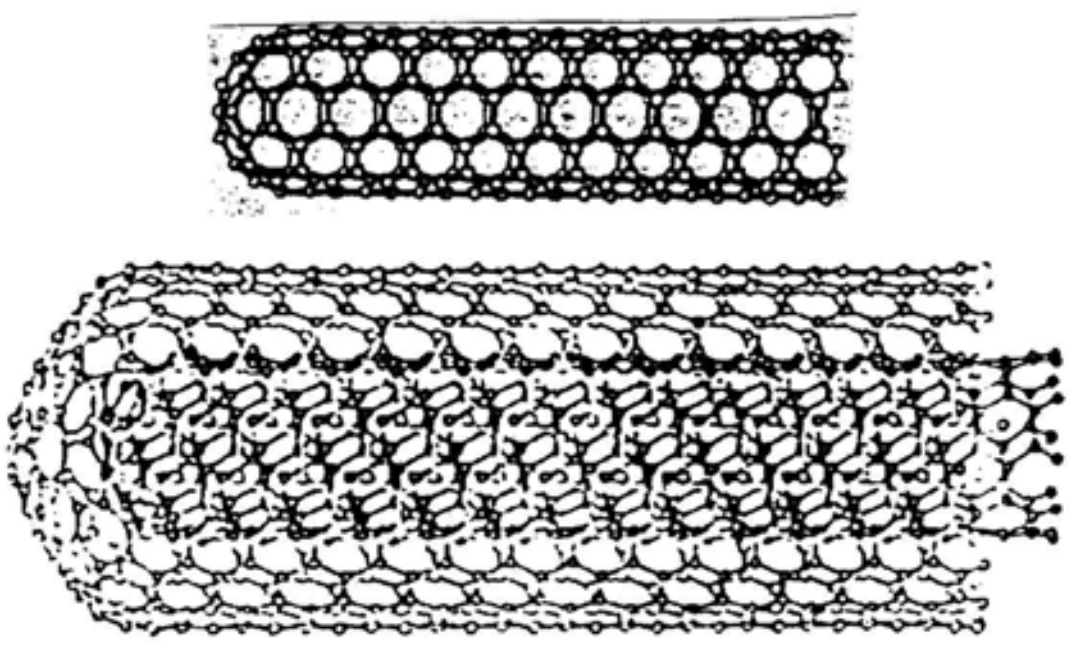 | Single-Walled CNT<br>Multi-Walled CNT |
| Carbon nanofoams | Carbon clusters with an average diameter ~6-9nm | |

**Fig. 1**